\documentclass[iop]{emulateapj}

\def\lapp{\ifmmode\stackrel{<}{_{\sim}}\else$\stackrel{<}{_{\sim}}$\fi}
\def\gapp{\ifmmode\stackrel{>}{_{\sim}}\else$\stackrel{>}{_{\sim}}$\fi}
\usepackage{multirow}

\begin{document}

\title{SPECTRAL AND TIMING PROPERTIES OF THE MAGNETAR CXOU~J164710.2$-$455216}

\author{
Hongjun An,
Victoria M. Kaspi\altaffilmark{1},
Robert Archibald, and \\
Andrew Cumming \\
{\small Department of Physics, McGill University, Rutherford Physics Building, 3600 University Street, Montreal, Quebec,
H3A 2T8, Canada}
}

\altaffiltext{1}{Lorne Trottier Chair; Canada Research Chair}

\begin{abstract}
We report on spectral and timing properties of the magnetar CXOU~J164710.2$-$455216 in the
massive star cluster Westerlund 1. Using 11 archival observations obtained with {\em Chandra} and
{\em XMM-Newton} over approximately 1000 days after the source's 2006 outburst,
we study the flux and spectral evolution of the source. We show that
the hardness of the source, as quantified by hardness ratio, blackbody temperature or power-law
photon index, shows a clear correlation with the 2--10 keV absorption-corrected flux and
that the power-law component flux decayed faster than the blackbody component
for the first $\sim$100 days. We also measure the timing properties of the source by analyzing
data spanning approximately
2500 days. The measured period and period derivative are 10.610644(17) \nolinebreak s (MJD 53999.06) and
$<4 \times 10^{-13}$ s s\textsuperscript{-1} (90\% confidence) which imply that the spin-inferred
dipolar magnetic field of the source is less than $7\times 10^{13}\ \rm G$. This is
significantly smaller than was suggested previously. We find evidence for a second flux
increase, suggesting a second outburst between MJDs 55068 and 55832.
Finally, based on a crustal cooling model, we find that the source's cooling curve can be
reproduced if we assume that the energy was deposited in the outer crust and that the temperature
profile of the star right after the 2006 outburst was relatively independent of density.

\end{abstract}


\section{Introduction}
Neutron stars with ultrastrong magnetic fields, so-called magnetars,
can emit radiation which is orders of magnitude stronger than their rotation power.
These objects are prone to X-ray outbursts in which the flux rises rapidly then decays
over days to months \citep[see][for reviews]{wt06, m08, re11}.
Such outbursts are often accompanied by short X-ray and soft
gamma-ray bursts. The origins of outbursts are still debated. It has been suggested that
a sudden or gradual internal heat release possibly caused by crustal
cracking, heats the star from within, and induces observed temperature and
luminosity increases \citep{td95, td96, pp11}. It has also been suggested
that crustal motion can generate twists in the external magnetic fields \citep{tlk02}.
The twisted magnetic fields induce currents in the magnetosphere, which return to the stellar
surface and heat it.

The flux relaxation following an outburst can be explained as passive
cooling of the hot star \citep[][see also Cumming et~al. 2012, in preparation]{let02, pr12},
or as ``untwisting'' of the external magnetic fields \citep{b09}.
In either case, flux and spectral hardness are expected to be correlated \citep{tlk02, l03, og07}.
The correlation has been observed in many magnetars
\citep[e.g., SGR 1806$-$20, 1E 2259$+$586, 1E 1547$-$5408;][]{wkf+07, zkd+08, sk11}
although it is not as clearly seen in some sources
\citep[e.g., SGR~1900$+$14, SGR~1627$-$41;][]{tem+07, akt+12}. 

\newcommand{\markv}{\tablenotemark{a}}
\newcommand{\marku}{\tablenotemark{b}}
\newcommand{\marky}{\tablenotemark{c}}
\newcommand{\markw}{\tablenotemark{d}}
\newcommand{\markt}{\tablenotemark{e}}
\begin{table*}[t]
\vspace{-0.2in}
\begin{center}
\caption{Summary of observations and spectral fit results
\label{ta:obs}}
\scriptsize{
\begin{tabular}{ccccccccccc} \hline\hline
\# & Date   & Observatory & Mode\markv & ID	&  $kT$	& $\Gamma$	& BB radius\marku& Flux\marky 	& PL Flux\markw & $\Delta T$\markt \\ 
  & MJD     &		  & 	&		  & (keV)&		&(km) 	 	& 	        &      	     & (s)	\\ \hline
1 & 53513.0 & {\em Chandra}& TE	 & 6283       & 0.51(2)&  $\cdots$	&0.50(5) & 0.030(2)    & $\cdots$  & 3.24 \\
2 & 53539.9 & {\em Chandra}& TE	 & 5411       & 0.53(14)&3.33(55)	&0.25(13)& 0.028(11)  & 0.02(1)   & 3.24 \\
3 & 53995.1 & {\em XMM}	& FW/FW  & 0404340101 & 0.59(6)& 3.86(22)	&0.18(4) & 0.025(4)    & 0.016(3)    & 2.6/0.073\\
4 & 54000.7 & {\em XMM}	& FW/FW  & 0311792001 & 0.70(1) & 2.90(6)	&1.44(5)& 3.00(7)   & 1.62(6)   & 2.6/0.073\\
5 & 54005.4 & {\em Chandra}& CC	 & 6724       & 0.60(1) & 2.46(13)	&2.22(9)& 2.52(9)   & 0.99(7)   & 0.00285\\
6 & 54010.1 & {\em Chandra}& CC	 & 6725       & 0.60(1) & 2.69(12) 	&2.08(8)& 2.08(7)   & 0.74(6)  	& 0.00285 \\
7 & 54017.4 & {\em Chandra}& CC	 & 6726       & 0.61(1) & 2.92(13) 	&1.97(6)& 1.84(6)   & 0.53(5)  	& 0.00285 \\
8 & 54036.4 & {\em Chandra}& CC	 & 8455       & 0.58(1) & 2.78(17) 	&1.97(9)& 1.42(6)   & 0.41(5)  	& 0.00285 \\
9 & 54133.9 & {\em Chandra}& CC	 & 8506       & 0.56(1) & 2.86(18) 	&1.83(9)& 0.95(4)   & 0.29(3)  	& 0.00285 \\
10& 54148.5 & {\em XMM}	& SW/LW  & 0410580601 & 0.58(1) & 3.20(11) 	&1.39(5)& 0.78(2)   & 0.27(2)  	& 0.3/0.048 \\
11& 54331.6 & {\em XMM}	& SW/LW  & 0505290201 & 0.58(1) & 3.42(10) 	&0.99(4)& 0.42(2)   & 0.17(1)	& 0.3/0.048 \\
12& 54511.5 & {\em XMM}	& SW/LW  & 0505290301 & 0.54(2) & 3.40(14) 	&0.93(7)& 0.26(1)   & 0.11(1)  	& 0.3/0.048 \\
13& 54698.7 & {\em XMM}	& SW/LW  & 0555350101 & 0.57(2) & 3.76(15) 	&0.62(4)& 0.16(1)   & 0.068(8) 	& 0.3/0.048 \\
14& 55067.6 & {\em XMM}	& SW/LW  & 0604380101 & 0.53(2) & 3.68(15) 	&0.56(4)& 0.087(5)  & 0.036(4) 	& 0.3/0.048 \\ 
15& 55832.0 &{\em XMM}	& SW/LW  & 0679380501 & 0.73(1)& 2.89(11)	&0.67(3) & 0.54(2)  & 0.18(2)	& 0.3/0.048 \\
16& 55857.8 & {\em Chandra}& TE subarray& 14360	 & 0.62(2)& 1.50(47)	&0.87(8) & 0.44(4)  & 0.18(2)	& 0.44\\ \hline
\end{tabular}}
\end{center}
\vspace{-0.1in}
\footnotesize{{\bf Notes.} The outburst occurred on 53999.05659 (MJD).
Obs.~3 (MJD 53995.1) was used to set the quiescent level for the flux evolution
(Fig.~\ref{fig:coolcurve}). $N_{\rm H}$ was obtained by simultaneously fitting Obs. 4--14, and was fixed for fitting
the others. Fits are conducted in the 0.5--10 keV band, and uncertainties are at the 1-$\sigma$ confidence level. \\}
$^{\rm a}${ MOS1,2/PN for the {\em XMM-Newton} observations.
TE: Timed Exposure, CC: Continuous clocking, FW: Full Window, LW: Large Window, SW: Small Window.}\\
$^{\rm b}${ Blackbody radius. Results of {\ttfamily tbabs(bbodyrad+pow)} fitting in {\ttfamily XSPEC} for an assumed distance of 5 kpc.}\\
$^{\rm c}${ Absorption-corrected flux in the 2--10 keV band in units of $10^{-11}$ erg cm\textsuperscript{$-$2} s\textsuperscript{$-$1}.}\\
$^{\rm d}${ Absorption-corrected power-law flux in the 2--10 keV band in units of $10^{-11}$ erg cm\textsuperscript{$-$2} s\textsuperscript{$-$1}.}\\
$^{\rm e}${ Time resolution.} \\
\vspace{-0.1in}
\end{table*}

The high magnetic fields of magnetars have been inferred both from spin-down rates, assuming the
standard dipole braking relation $B \equiv 3.2 \times 10^{19}(P \dot P)^{1/2}\ \rm G$ \citep{mc77},
as well as from indirect arguments regarding radiative behavior and field decay
\citep{td95, td96}. Indeed the typical magnetic field for an object classified as a magnetar on
the basis of radiative behavior is $10^{14}-10^{15}$~G.\footnote{See the online magnetar catalog for
a compilation of known magnetar
properties, http://www.physics.mcgill.ca/$\sim$pulsar/magnetar/main.html}
However recently a small handful of objects have been reported to have fields
below this range, overlapping with those of apparently ordinary radio pulsars
\citep[SGR~0418$+$5729, Swift~J1822.3$-$1606;][]{ret+10, lsk+11, rie+12, snl+12}.
Although the $B$-distribution of radiatively classified magnetars still remains significantly
higher than that of radio pulsars \citep{lsk+11}, the apparently low-$B$
magnetars are puzzling and are suggestive of higher order multipolar structure or
strong internal toroidal fields, for which the dipole component is deceptively low \citep{tzp+11}.

CXOU~J164710.2$-$455216 was discovered on 1998 June 15 with
the {\em Chandra X-Ray Observatory} \citep{mcc+06}. It is located at
R.A. = 16\textsuperscript{h}47\textsuperscript{m}10\textsuperscript{s}.20,
Decl. = $-$45$^\circ$52$'$17$''$.05 \citep[J2000.0,][]{spz06}
and is estimated to be $2.5 - 5$ kpc away, in the massive star cluster Westerlund 1 \citep{cnc+05}.
A short 20-ms burst was detected with the {\em Swift} Burst Alert Telescope (BAT)
on 2006 Sept. 21 \citep[MJD 53999.05659,][]{kbc+06}. The source was subsequently
observed with multiple X-ray observatories
including {\em Swift}, {\em XMM-Newton}, {\em Chandra} and {\em Suzaku}.
\citet{icd+07} and \citet{wkg+11} reported on the data obtained after the 2006 outburst out
to approximately 100--200 days and measured spectral and timing properties, where
they concluded that there is no clear evidence of significant spectral variability.
The pulsar's spin period is $\sim 10.6$ s \citep{mcc+06} and
the spin-down rate has been reported to be $0.8-1.3\times10^{-12}$ s s\textsuperscript{$-$1}
\citep{icd+07, wkg+11}. These previously reported spin parameters imply an inferred
surface dipolar magnetic-field strength of $B=0.95-1.2 \times 10^{14}$ G.

Here, we report on the analysis of {\em Chandra} and {\em XMM-Newton} archival
data which cover a longer time period ($\sim$2500 days).
We study the long-term spectral evolution of the source after
its outburst and report on the pulsar's timing behavior during this period.

In Section~\ref{sec:obs}, we summarize the observations we used for our study. We then describe
our data analysis and results in Section~\ref{sec:ana},
discuss the results in Section~\ref{sec:disc}, and
finally present our conclusions in Section~\ref{sec:concl}.

\section{Observations}
\label{sec:obs}
Table~\ref{ta:obs} presents the data we used for our analysis. In total, we analyzed
16 observations  obtained with the {\em Chandra} and {\em XMM-Newton} telescopes over the
course of $\sim$2500 days.
For the {\em Chandra} observations, we reprocessed the standard pipeline output using
{\ttfamily chandra\_repro} of CIAO 4.4\footnote{http://cxc.harvard.edu/ciao4.4/index.html}
along with CALDB 4.4.7 to produce the
``level 2'' event lists using the newest software and calibration
updates.\footnote{http://cxc.harvard.edu/ciao/threads/createL2/}
For the {\em XMM-Newton} data, we processed the Observation Data Files (ODF) with {\ttfamily epproc} and
{\ttfamily emproc} and then applied the standard filtering procedure
(e.g., flare rejection and pattern selection) of
Science Analysis System (SAS) version 11.0.0.\footnote{http://xmm.esac.esa.int/sas/}
Some observations in the Table were already analyzed by other authors
\citep[][]{mcc+06, mgc+07, icd+07, wkg+11}.
However, we re-analyzed those using the above procedure for consistency.

\section{Data Analysis and Results}
\label{sec:ana}
\subsection{Imaging Analysis}
\label{imageana}
We detected the source with {\ttfamily wavdetect} for the {\em Chandra} TE mode observations and
with {\ttfamily edetect\_chain} for the {\em XMM-Newton} observations. The source positions we found
are all consistent with one another and with the known source location within 90\% uncertainties of the
position measurements (0$''$.6 for
{\em Chandra}\footnote{http://cxc.harvard.edu/cal/ASPECT/celmon/} and 2$''$ for
{\em XMM-Newton}\footnote{http://xmm2.esac.esa.int/external/xmm\_sw\_cal/calib}).

The relatively large point-spread function of {\em XMM-Newton} was a concern since the source
is located in a star cluster, and any other object within $\sim 30''$ could in principle contaminate the source
spectrum. With two {\em Chandra} TE observations (IDs 6283 and 14360), we searched for X-ray sources
within a radius of 30$''$ centered at CXOU~J164710.2$-$455216 but found none.
We also checked whether our source is consistent with being a point source using the {\ttfamily eradial} tool for
{\em XMM-Newton} observations and Chandra Ray Tracer\footnote{http://cxc.harvard.edu/chart/} and the
{\ttfamily MARX}\footnote{http://space.mit.edu/CXC/MARX} tools for the {\em Chandra} observations.
The data were consistent with the target being a point source in all the observations except for
the first {\em XMM-Newton} observation (ID: 0311792001), where the radial profile of the source was
distorted due to pile-up.
Therefore, we conclude that there is no significant contamination from unresolved sources in the
{\em XMM-Newton} observations and that the source exhibits no detectable extended emission.

\subsection{Spectral Analysis}
\label{spectrumana}
We used all the data listed in Table~\ref{ta:obs} for our spectral analysis. For the {\em Chandra}
observations, we extracted the source events using a box of dimension 2$''$ along the 1-D events
and 5$''$ in the orthogonal direction for the CC-mode observations, and a circle with radius
$2''$ for the TE-mode observations.
Backgrounds were obtained in two rectangular regions with size of $10'' \times 5''$ from each side of
the source region and an annular region with radii of 5$''$ and 10$''$ centered at the source for
the CC-mode and the TE-mode observations, respectively.
Then we produced spectra using
the {\ttfamily specextract} tool of CIAO 4.4 with CALDB 4.4.7 and grouped them to have a minimum of
20 counts per bin for further analysis.

For the {\em XMM-Newton} data, we extracted source spectra from circular regions having radius of
16$''$ and background spectra from source-free regions on the same chip.
Corresponding response files were produced using the {\ttfamily rmfgen} and the {\ttfamily arfgen} tasks
of SAS 11.0.0. Each spectrum was then grouped to have a minimum of 20 counts per bin.

The first {\em XMM-Newton} observation was mildly piled up, so we excluded the central 
$\sim$5$''$ in each of the PN and the two MOS detectors to minimize the pile-up effect.
After removing the central regions, we checked if pile-up was still
significant using the {\ttfamily epatplot} tool of SAS.
With the removal, the measured event pattern distributions showed good
agreement with the expected ones.
We also checked if the spectral parameters changed significantly when we removed central regions of
different sizes. The power-law index changed smoothly by approximately 3\% when we varied the removal
radius from 0$''$ to 9$''$ for the PN data, but it changed abruptly (softened by 10\%) for removal
regions between 0$''$ and 3$''$.5 and then stayed constant for the MOS data.
Therefore, we ignored the central 5$''$ for both the PN and
the MOS data to ensure that pile-up would not distort the spectrum of the source in this one observation.

We fit the 0.5--10 keV spectra with three models: an absorbed power law plus blackbody, an absorbed double
blackbody and an absorbed blackbody ({\ttfamily tbabs*(power + bbody)}, {\ttfamily tbabs*(bbody + bbody)} and
{\ttfamily tbabs*bbody} in {\ttfamily XSPEC 12.7.0}).\footnote{http://heasarc.gsfc.nasa.gov/xanadu/xspec/}
We fit all the data in which the source
flux decreased monotonically after the 2006 outburst (Obs.~4--14; see Table~\ref{ta:obs}) simultaneously with
a common hydrogen column density ($N_{\rm H}$).

The single blackbody fit ($\chi^2/DoF=9117.42/7062$) and double blackbody fit ($\chi^2/DoF=7353.82/7040$)
were not acceptable as they showed systematic trends in the low energy ($< 1$ keV) and the high energy bands
($> 8$ keV).
However, the blackbody plus power-law fit described the observed spectra well ($\chi^2/DoF=6988.04/7040$).
The $N_{\rm H}$ we obtained from the fit is $2.39(5) \times 10^{22}$ cm\textsuperscript{$-$2},
which is different from the values that \citet[][1.44 $\times 10^{22}$ cm\textsuperscript{$-$2}]{mgc+07}
and \citet[][1.9 $\times 10^{22}$ cm\textsuperscript{$-$2}]{icd+07} used. However, it is
consistent with that obtained by \citet{wkg+11} and that reported by
\citet[][2.1--3.0$\times 10^{22}$ cm\textsuperscript{$-$2}]{spz06}
on the basis of cluster extinction estimates of \citet{cnc+05} and the empirical relation of \citet{g75}.
We then fixed $N_{\rm H}$ and tried to fit the spectra of the other, fainter observations with the same models.
The results are summarized in Table~\ref{ta:obs}.

We assumed that the source was in (or near) quiescence for the first three observations (Obs.~1--3). The source
spectrum was well fitted with a single blackbody for Obs.~1 without requiring an additional component, although
a power law was not ruled out unambiguously.
We fit Obs.~2 and 3 with a power law;
a single blackbody fit was unacceptable. However, adding another component (blackbody) significantly improved the fit
(F-test probabilities of 0.048 and 0.00063 for Obs.~2 and 3, respectively).

\begin{figure}
\centering
\vspace{0.1 in}
\begin{tabular}{c}
\vspace{0.1 in}
\includegraphics[width=2.25 in, angle=90]{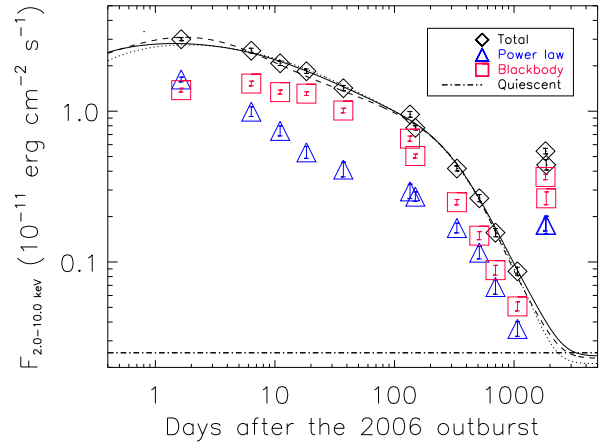} 
\end{tabular}
\vspace{-0.15 in}
\figcaption{Flux evolution of CXOU~J164710.2$-$455216 after the 2006 outburst and crustal cooling model fits.
Symbols are 2--10 keV absorption-corrected flux evolutions for total (diamonds), power-law (triangles) and
blackbody (squares) components, and lines are quiescent flux (dot-dashed, Obs. 3 in Table~\ref{ta:obs}) and
crustal cooling models for decay of the source flux
(see Section~\ref{sec:fluxevol} for details).
Solid line: $T_c=1.1\times10^8\ \rm K$, $B=8\times10^{13}\ \rm G$, $Q_{imp}=5$, initial temperature
changes from $1.37\times10^9\ \rm K$ at $9\times10^8\ \rm g\ cm^{-3}$ to
$4.5\times 10^8\ \rm K$ at $5\times10^{11}\ \rm g\ cm^{-3}$.
Dotted line: $T_c=9\times10^7\ \rm K$, $B=2\times10^{14}\ \rm G$, $Q_{imp}=1$,
initial temperature changes from $1.3\times10^9\ \rm K$ at $\rho=1\times10^9\ \rm g\ cm^{-3}$
to $4\times10^8\ \rm K$ at $\rho=5\times10^{11}\ \rm g\ cm^{-3}$.
Dashed line: $T_c=5.7\times10^7\ \rm K$, $B=2\times10^{15}\ \rm G$, $Q_{imp}=5$,
initial temperature is $T=1.5\times10^9\ \rm K$ for
$1.5\times10^9\ \rm g\ cm^{-3} < \rho < 1\times10^{11}\ \rm g\ cm^{-3}$.
Other parameters were held fixed: $M=1.4M_{\sun}$, $R=12\ \rm km$. Note that
the uncertainty for each data point is smaller than the symbol size.
\label{fig:coolcurve}
}
\end{figure}

Another flux increase was observed in Obs.~15, which might be due to a second outburst between MJDs
55068 and 55832 (see Fig.~\ref{fig:coolcurve}).
We were not able to fit the spectrum for Obs.~15 or 16 with single-component models. For Obs.~15, a
blackbody plus power law provided a better fit ($p=0.080$) than a double blackbody
($p=0.033$ for $kT_1=0.36(2)$ keV, $kT_2=0.82(2)$ keV). For Obs.~16, a double blackbody model fit the
data better with $kT_1=0.58(3)$ keV and $kT_2=1.33(35)$ keV although a
blackbody plus power law also provided an acceptable fit.
Since it seems unlikely that the blackbody temperature increased from $0.80$ keV to $1.33$ keV without
any accompanying flux increase, we report the blackbody plus power-law spectrum for Obs.~15 and 16.
The results of all our spectral fits are reported in Table~\ref{ta:obs}.
 
\begin{figure*}
\centering
\vspace{8mm}
\begin{tabular}{cc}
\vspace{12.00mm}
\includegraphics[width=2.27 in,angle=90]{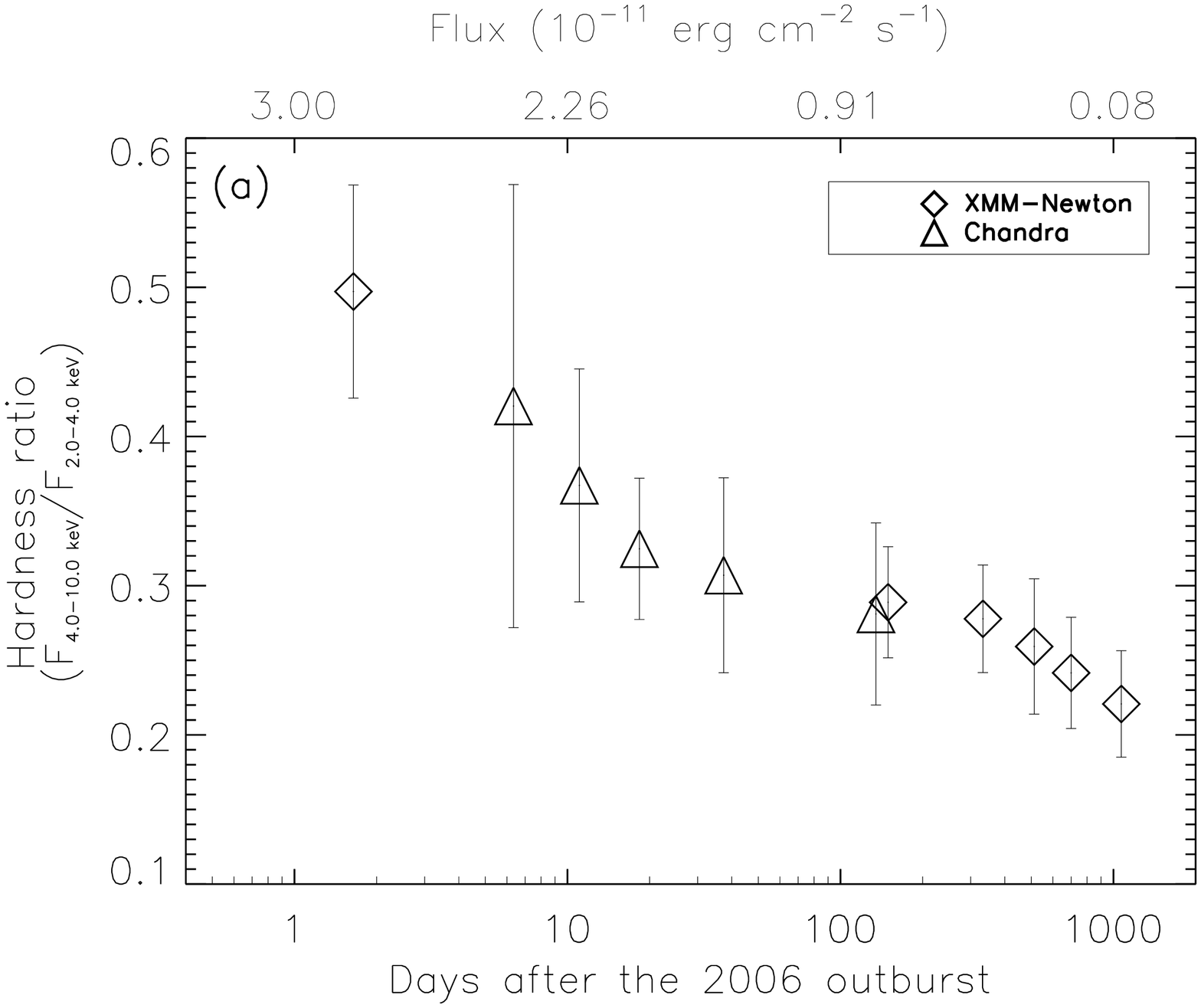} &
\includegraphics[width=2.27 in,angle=90]{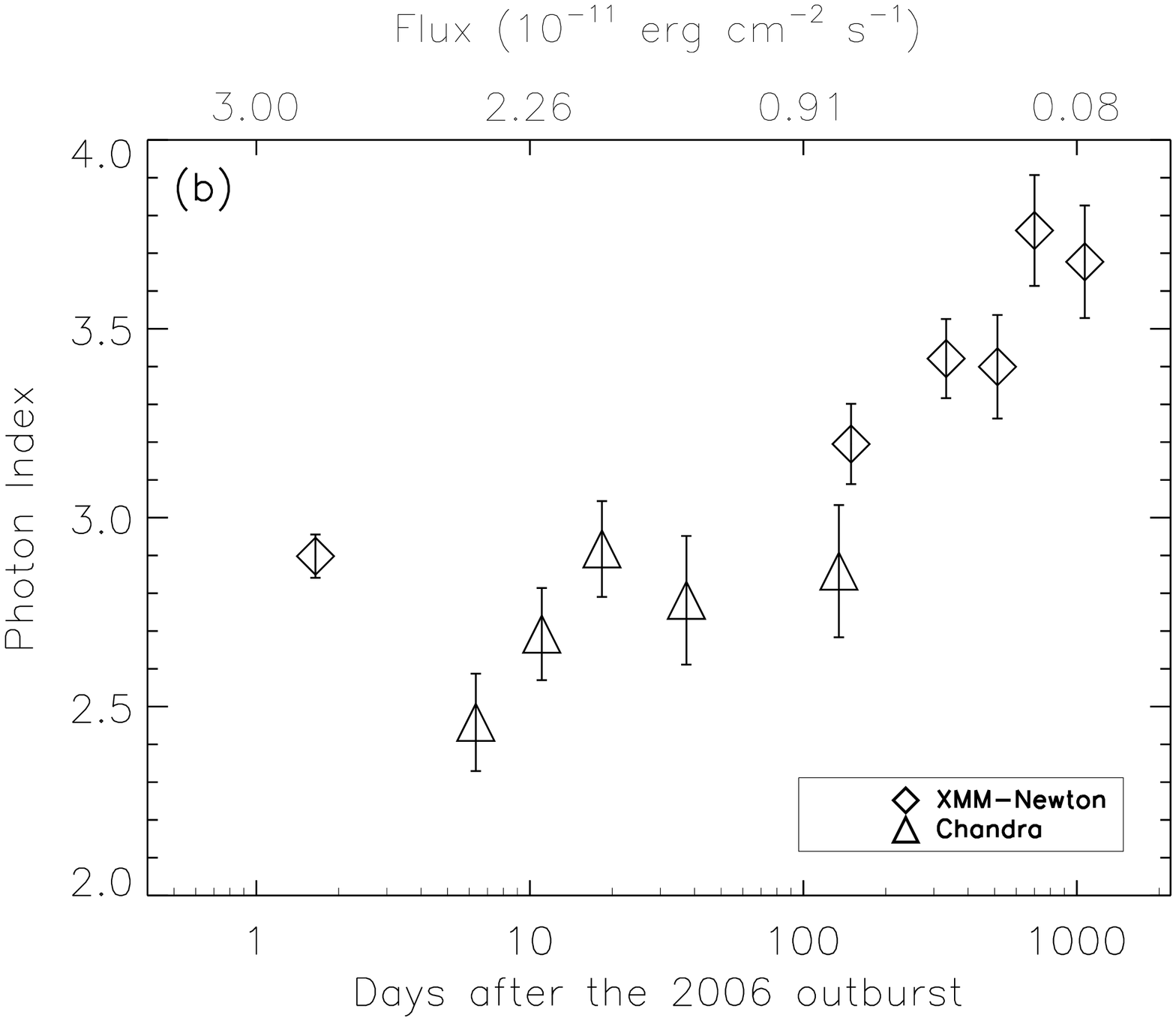} \\
\includegraphics[width=2.27 in,angle=90]{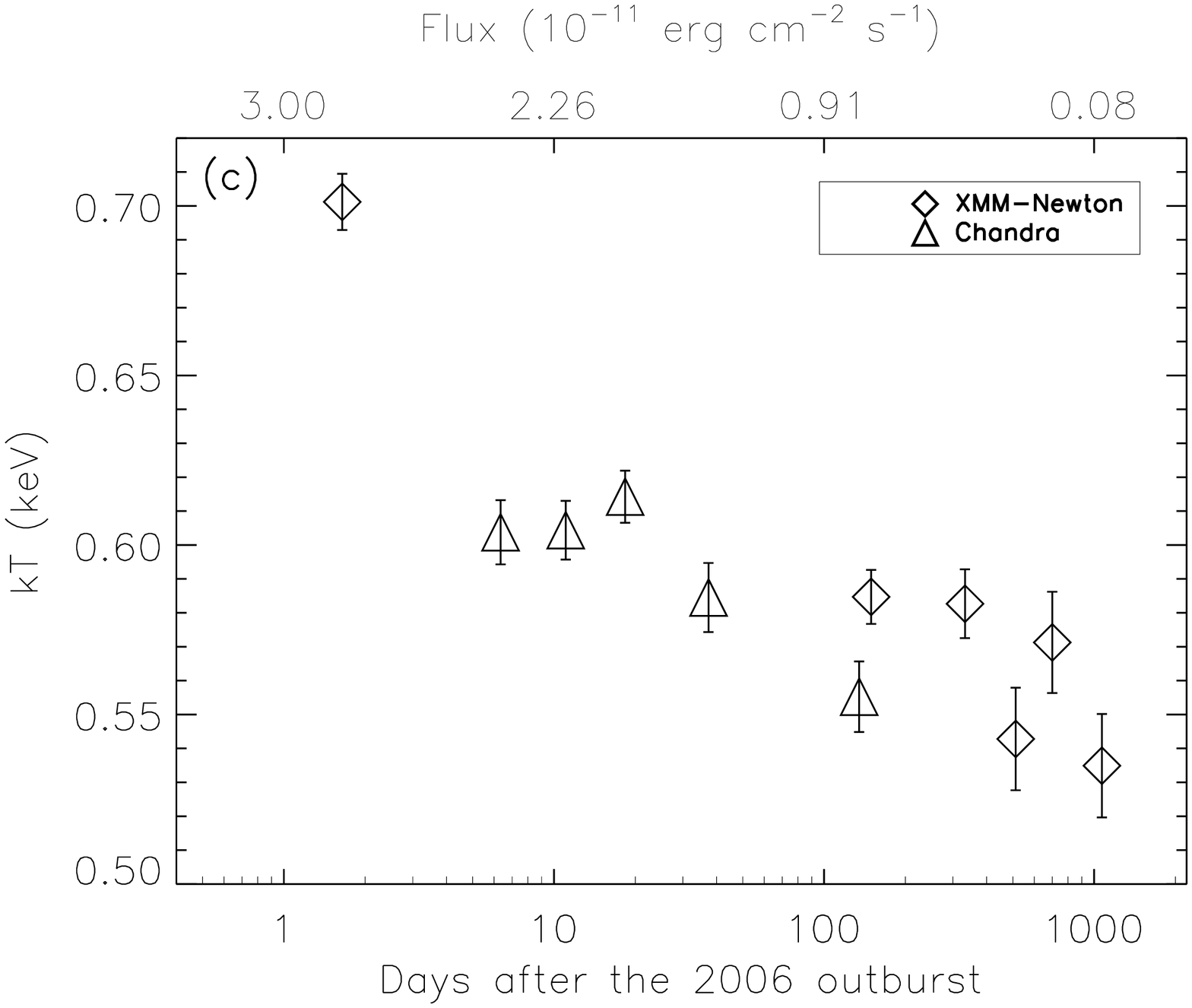} &
\includegraphics[width=2.27 in,angle=90]{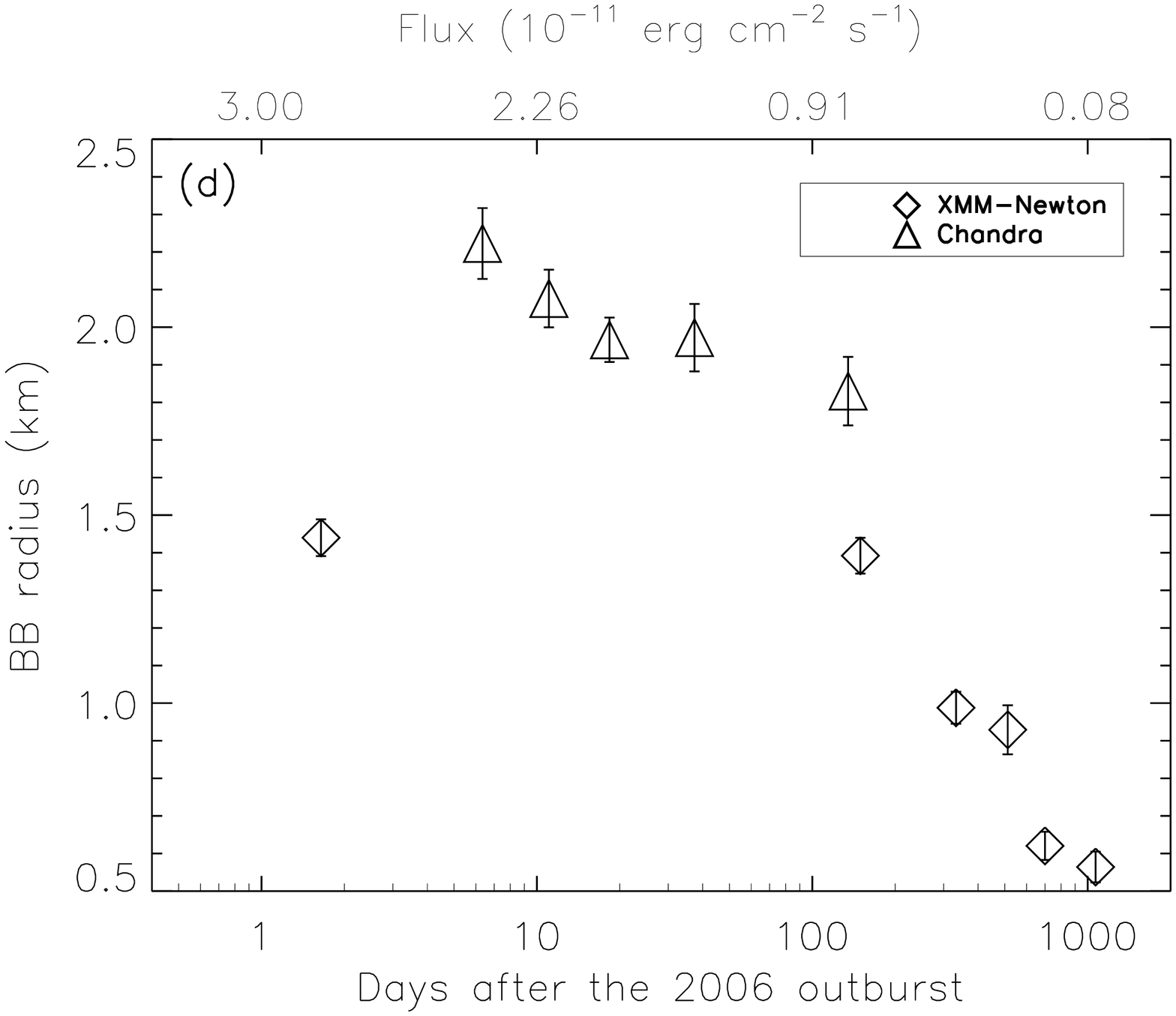}
\end{tabular}
\figcaption{The time evolution of various spectral properties.
The top axis shows the 2--10 keV absorption-corrected flux interpolated with
the double-exponential relaxation fit to the flux evolution (see section~\ref{specevol}).
The conversion between time and flux
was obtained from the double exponential fit result. Clear trends in time (flux) are visible.
{\em a}) Hardness ratio ($F_{4-10\ \rm keV}/F_{2-4\ \rm keV}$).
{\em b}) Power-law photon index.
{\em c}) Blackbody temperature.
{\em d}) Blackbody radius assuming a distance of 5 kpc.
\label{fig:coolpow}
}
\end{figure*}

We also fitted the spectra for Obs.~3--14 with an absorbed resonant cyclotron
scattering model, {\ttfamily tbabs*(atable\{RCS.mod\})} in {\ttfamily XSPEC} \citep[][]{rzt+08}. From the fit,
we obtained a smaller $N_{\rm H}$ ($2.06\pm0.02\times10^{22}\ \rm cm^{-2}$), as expected because of the strong
cutoff in the power-law spectrum at low energies. We were able to measure 2--10 keV absorption-corrected
fluxes using the model, and they agreed well with those of the blackbody
plus power-law fits described above.
However, the RCS model parameters were not well constrained.
Therefore, we mainly use the results of the blackbody plus power-law fit for discussion below.

\subsection{Spectral Evolution}
\label{specevol}
We plot the time evolution of the 2--10 keV absorption-corrected fluxes after the 2006
outburst in Figure~\ref{fig:coolcurve}.
We have attempted to fit the data out to $\sim$1000 days with a double
exponential, $F(t) = F_1e^{-(t - t_0)/\tau_1}+F_2e^{-(t - t_0)/\tau_2} + F_Q$ or a broken power
law, $F(t) = F_1(t - t_0)^{\alpha_1} + F_Q$ for $t_0<t<T_{break}$ and
$F(t)=F_2(t - t_0)^{\alpha_2} + F_Q$ for $T_{break}<t$, where $F_1$ and $F_2$ are determined so that
$F(t)$ is continuous at $t=T_{break}$.
In fitting, we set $t_0$ to be the onset of the outburst (MJD 53999.05659) and fixed the
quiescent flux ($F_Q$) to be the measured value of $2.5\times10^{-13}$
erg cm\textsuperscript{$-$2} s\textsuperscript{$-$1} (Obs.~3 in Table~\ref{ta:obs}).

Although the fit results with the simple analytical models were statistically unacceptable,
the double exponential function roughly described the relaxation with decay times of $18\pm2$ and $325\pm10$ days
($\chi^2$/dof=33.5/7), as did the broken power-law fit with decay indices of $-0.22\pm0.01$ and $-1.16\pm0.02$,
and $T_{break}$ of $106\pm5$ days ($\chi^2$/dof=64.8/7). Note that we include only statistical
uncertainties in fitting, although some systematic for example,
the sudden decrease in flux at $\sim$150 days in
Figure~\ref{fig:coolcurve} may indicate a cross-calibration issue between
{\em Chandra} and {\em XMM-Newton}. Indeed, adding a 6\% systematic uncertainty to the fluxes
made the double exponential fit acceptable (decay times of $19\pm4$ and $331\pm15$ days, $\chi^2$/dof=12.3/7).

\newcommand{\markh}{\tablenotemark{a}}
\newcommand{\marki}{\tablenotemark{b}}
\begin{table*}[t]
\vspace{-0.05in}
\begin{center}
\caption{Timing properties of CXOU~J164710.2$-$455216
\label{ta:timing}}
\scriptsize{
\begin{tabular}{cccccc} \hline\hline
			& Period 	& Period Derivative		&	Epoch	& $B$\markh	& Comment \\  
 			& (s)	 	& (s s\textsuperscript{$-$1})	&	(MJD)	& ($10^{14}$ G)	& 	\\  \hline   
\citet{icd+07} 		& 10.6106549(2)	& $9.2(4)\times10^{-13}$ 	&53999.0	& 1.00		&$\cdots$\\ \hline
\multirow{2}{*}{\citet{wkg+11}}	& 10.6106567(1)	& $8.3(2)\times10^{-13}$&\multirow{2}{*}{54008.0}& 0.95	&$\cdots$\\
                      	& 10.6106567(2)	& $1.3(1)\times10^{-12}$ 	&		& 1.19		& with cubic term\\ \hline
This work		& 10.610644(17)	& $<4\times10^{-13}$\marki 	&53999.1	& $<$0.7\marki	&$\cdots$\\
\hline
\end{tabular}}
\end{center}
\vspace{-0.1in}
\footnotesize{{\bf Notes.}
}
$^{\rm a}${ Spin-inferred dipolar magnetic field.}\\
$^{\rm b}${ 90\% upper limit.} \\
\end{table*}

We also plot the time evolutions of the spectral properties for the same data in Figure~\ref{fig:coolpow}
in order to look for any correlation between spectral hardness and flux,
as seen in other magnetars \citep[e.g.,][]{roz+05, wkf+07, cri+07, tgd+08, zkd+08, sk11}.
Clear trends are visible in the hardness
ratio, power-law photon index and blackbody temperature; the spectrum clearly softened as the flux decreased.

For the first $\sim$100 days, we were able to reproduce the same trends as those of \citet{wkg+11}
for the spectral parameters, though with small offsets within uncertainties
in the power-law index and the blackbody temperature. The offsets could be due to the fact that they
used a different CALDB or analysis software version and/or included a {\em Suzaku} observation.

\subsection{Timing Analysis}
\label{timingana}

Timing studies have been done by \citet{icd+07} and \citet{wkg+11} using
phase-coherent analyses with data obtained for approximately 100--200 days after the 2006 outburst.
Since the separation between observations 10 and 11 was large, phase connection was lost.
Instead, we searched for the best pulse period in
each observation to find the average spin-down rate. To do this, we employed the  $H$-test \citep{dsr+89} because
it requires no advance knowledge of the pulse profile at any epoch, and because this source's pulse
profile is known to change \citep[][]{icd+07, wkg+11}.

The pulse period of the source is $\sim$10.6 s. The time resolutions of the observations are all smaller than $\sim$5 s
so could be used in our analysis (see Table~\ref{ta:obs}).
Thus, the baseline useful for timing spans approximately 2500 days.
We extracted source events from the regions described
in Section~\ref{spectrumana}. Then we selected events having energies in the range 0.5--8 keV for the
analysis. Each observation was then corrected to the barycenter using the {\ttfamily axbary} tool of CIAO for
the {\em Chandra} data and the {\ttfamily barycen} tool of SAS for the {\em XMM-Newton} data.

For each observation, we calculated the sum of the Fourier powers of $n$ harmonics,
$Z_n^2(P)$ \citep{bbb+83} for $n$=1--20 at test periods in the range
10.6050--10.6160 s with step size $\Delta P=10^{-5}$ s
and found $n$ and $P$ which maximized $H=Z_n^2(P)-4n+4$ \citep{dsr+89} for each observation.
We folded the events at
the best period to obtain a pulse profile. We used simulations to determine the uncertainty on the best period.
To do this, we used the measured pulse profiles (which had 40 bins) to
produce fake event lists containing the periodicities. Fake event lists were made by first
taking an observed pulse profile, and adding Poisson noise as well as a random phase offset to it.
Then, for each bin in the
profile, we took the total number of counts, assigned each one a random arrival time in that
phase bin and rebinned the arrival times to the time resolution of the detector.
In this way, we constructed a simulated event list. For each such list, we then
calculated the $H$ statistics for each $P$ in the range given above, and for all $n$'s in the range
of 1--20. We then found $P$ and $n$ that maximized $H$, and took the standard deviation of
the periods as the period uncertainty. The uncertainties we obtained in this way agree with those calculated using
the method of \citet{rem02} within a factor of two, but typically are slightly larger.
To run the simulations for observations with coarse time resolution where the measured pulse profiles were
not well determined, we assumed the XMM PN profile.

\begin{figure}
\vspace{0.1in}
\centering
\includegraphics[width=2.25 in,angle=90]{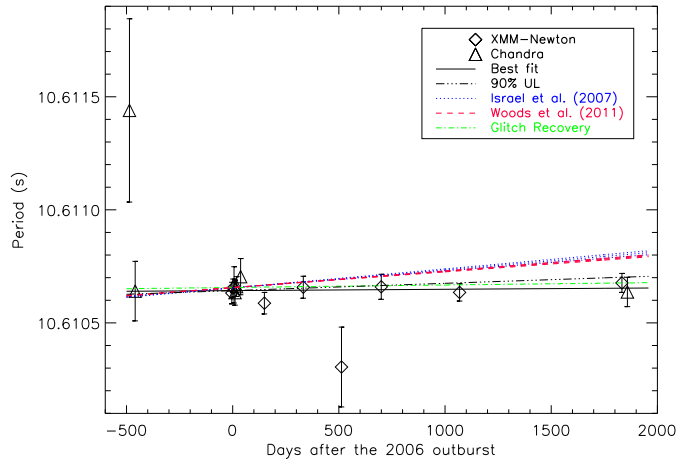}
\figcaption{The evolution of pulse period after the 2006 outburst. Periods are measured using the
$H$-test \citep{dsr+89} and uncertainties are estimated using simulations (see text).
The solid line is the best fit to the data, the triple dot-dashed line is the 90\% upper limit for the
best fit, and the dotted and the dashed lines are the expected
evolution of the period in time for the previously reported period derivative with 90\% confidence
interval by \citet[][]{icd+07} and \citet{wkg+11} with no cubic term, respectively. Also plotted
in the dot-dashed line is the solution of \citet{wkg+11} assuming glitch recovery following the 2006 outburst.
\label{fig:PPdot}
}
\end{figure}

In Figure~\ref{fig:PPdot}, we show the best period we obtained for each observation. For the {\em XMM-Newton}
observations, we took the variance-weighted average period for measurements done with the three instruments
(MOS1, MOS2 and PN).
We then fit the periods with a linear function ($P(t)=P_0+\dot P t$) to obtain the best period at the time of
outburst, $P_0$ and the spin-down rate. The fit results were $P_0=10.610644\pm0.000017$ s
and $\dot P=0.7\pm2.4\times10^{-13}$ s s\textsuperscript{-1}. The period derivative we measured is consistent with
zero, so we report the 90\% upper limit, $4 \times 10^{-13}$ s s\textsuperscript{-1}.
This is significantly smaller than what
\citet{icd+07} and \citet{wkg+11} reported (see Table~\ref{ta:timing}, and Section~\ref{sec:timing} for discussion).

We also searched for short-term aperiodic variability of the source in the 0.5--8 keV band.
For the {\em XMM-Newton} data,
we produced light curves using {\ttfamily evselect} and corrected them for detection efficiency
with the {\ttfamily epiclccorr} tool of SAS. We binned the light curves at 30 to 2000 s intervals
(ensuring at least 20 events per bin) depending on the source flux
since it was declining. We then fit the light curves to a constant, and calculated
$\chi^2$ and the null hypothesis probability.
For the {\em Chandra} CC-mode observations, it is difficult to correct the detection efficiency
because they lack information on one of the two dimensions.
Therefore, we assumed a uniform detection efficiency for the data taken with CC-mode.
We then performed the $\chi^2$ test and the Gregory--Loredo test \citep[using the
{\ttfamily glvary} tool of CIAO;][]{gl92} for all the {\em Chandra} observations in Table~\ref{ta:obs}.
In all cases, the observed light curves were consistent with being constant.

\section{Discussion}
\label{sec:disc}
We have measured the spectral and timing properties of CXOU~J164710.2$-$455216 for three years after its
2006 outburst.
We observe a clear correlation between spectral hardness and flux post-burst. 
In our timing analysis, we find a significantly smaller period derivative
which implies a spin-inferred dipolar magnetic field of $<7\times 10^{13}\ \rm G$ (90\% confidence).
This is significantly lower than was previously reported.
We find evidence that a second outburst occurred between MJDs 55068 and 55832 based on a flux increase at
MJD 55832 (Obs. 15).
Next we discuss our findings in relation to previous
studies of this source, and in the context of the magnetar model.

\subsection{Correlation between Spectral Hardness and Flux}
The spectral evolution of the source after the 2006 outburst has been discussed in previous studies \citep{icd+07, wkg+11}.
Based on 100--200 days of observations, \citet{icd+07} and \citet{wkg+11} found no
clear evidence that the spectral shape ($kT$ and $\Gamma$) had evolved. Our analysis results for
the same period qualitatively agree with theirs (see the first seven data points of the $\Gamma$ and $kT$
evolution plots in Figure~\ref{fig:coolpow}). 

However, the power-law photon index may not be a reliable measure of spectral hardness because a power law
may not be the true spectrum, especially at low energies, as noted by many authors
\citep[e.g.,][]{lg06, og07, ft07}.
Several authors showed that hardness ratio correlates with flux more strongly \citep[e.g.,][]{og07, zkd+08}.
For this reason, we plot the hardness ratio ($F_{4-10\ \rm keV}/F_{2-4\ \rm keV}$) as a function of 2--10 keV flux in
Figure~\ref{fig:coolpow}. The hardness ratio showed a clear correlation with flux even for
the first $\sim$100--200 days after the 2006 outburst. This proves that the spectral shape changed
during that period. Moreover, given the longer baseline of the observations, we now discern a clear
trend of photon index and $kT$ with flux, in agreement with spectral softening as the source relaxes.

We note however that the first data point obtained 1.7 days after the outburst epoch lies outside the trend
for $\Gamma$. The source was hottest ($\sim$0.7 keV) at that epoch
but the power-law component was not hardest.
For a crustal event \citep[see][]{mgc+07}, this may be explained as follows.
When a crustal event occurs, the crustal fracture may implant twists in the magnetosphere.
The twisted fields then induce currents in the magnetosphere \citep{tlk02, bt07}.
Perhaps the magnetospheric plasma was not yet fully activated at the epoch of the first post-outburst observation.
This would be consistent with what we infer from the change in the blackbody radius; the
magnetospheric model predicts that the X-ray emitting area decreases monotonically \citep{b10}, whereas the
data show an increase in the beginning (see blackbody radius plot in Figure~\ref{fig:coolpow}).

We note that the flux evolution after Obs.~15 showed a similar trend as during
early times after the 2006 outburst: the power-law spectrum hardened while the blackbody temperature
decreased and the radius increased from $0.67\pm0.03$ km (Obs.~15) to $0.87\pm0.08$ km (Obs.~16).

\subsection{Flux Evolution}
\label{sec:fluxevol}
\citet{icd+07} suggested that the
2--10 keV absorption-corrected flux relaxation of the source followed a power-law decay with index of
$-0.28 \pm 0.05$ for approximately 120 days, based on {\em Swift} observations.
\citet{wkg+11} obtained a power-law decay with index of $-0.306\pm0.005$ for the flux
evolution for the first $\sim$200 days based on {\em Chandra}, {\em XMM-Newton} and {\em Suzaku}
observations.
We find that either a double exponential or a broken power-law function describes the long-term
cooling trend of the source, although a double exponential provides a significantly better fit.
For the first 100--200 days, the cooling trend can be described by a power-law decay with index
of $-0.22\pm0.01$.
This is roughly consistent with the previous measurements of \citet{icd+07} and \citet{wkg+11}.
Note that the two data points between 100 days and 200 days are attributed to the second power-law
decay component in our fit while \citet{wkg+11} used a single power law out to $\sim 200$ days. However,
we find that the flux evolution at later times changed significantly as is seen in Figure~\ref{fig:coolcurve}.
This behavior is observed in some magnetars' post-outburst relaxation \citep[e.g.,][]{wkt+04, lsk+11, akt+12}
and can be qualitatively explained by crustal cooling models \citep[e.g.,][]{let02} and/or the untwisting
magnetic fields model \citep{b09}.

Further, \citet{icd+07} argued that the power-law component flux decayed more rapidly than the blackbody
component did for the first 4 months.
They argued from this that hot spots on the surface sustained heat longer than the external source did.
However, \citet{wkg+11} found no evidence that the power-law spectral component declined
faster based on the stability of the power-law to blackbody flux ratio during cooling.
Our results are consistent with the former; the power-law spectral component
decayed faster than the blackbody component for the first $\sim$100 days.
This trend is clearly visible in Figure~\ref{fig:coolcurve}.
Also note that the power-law decay indices we measured are consistent with those of
\citet{icd+07} although the fit was not good for the blackbody component.

\citet{b09} proposed the ``untwisting'' magnetospheric model to explain transient cooling of magnetars.
\citet{b10} argues that the area of a hot spot shrinks in the ``untwisting'' model while it should not
in the crustal cooling case since heat is expected to diffuse. 
On the basis of observations of several magnetars' transient cooling,
\citet{b10} argues that magnetospheric untwisting plays a significant role
in the transient cooling. For the cooling of CXOU~J164710.2$-$455216 after its 2006 outburst, there
is evidence that the area of the hot spot increased for the first $\lesssim 10$ days and then decreased,
if we assume that the blackbody component in our spectral model reasonably represents the true
thermal emission from the source. In this scenario, the evolution of the blackbody radius and the photon index
imply that the relaxation is a combination of the crustal and the magnetospheric effects with
the dominant process being the crustal cooling at early times. Whether or not CXOU~J164710.2$-$455216 is
the only source that showed this behavior is not clear. An increase of the blackbody
radius at the very early stages of relaxation and subsequent decrease has been observed in other
magnetars \citep[e.g., SGR~0501$+$4516 and SGR~0418$+$5729; ][]{rit+09, eit+10}. However, sudden
hardening of the power-law component was not observed for SGR~0501$+$4516, and a power-law component was not
detected significantly for SGR~0418$+$5729.

With the current level of theory predictions and model developments, it is difficult to
explain the changes of all the observational properties such as flux, spectrum and blackbody area
simultaneously, and to know unambiguously whether the external (magnetospheric) or the internal (crustal)
effect dominates at any epoch. 
Nevertheless, here we consider a crustal cooling model in more detail, and show that it can reproduce the
observed flux decay.
In the model, we calculate the thermal relaxation of the crust after a rapid energy deposition by solving
the diffusion equation \citep[see Cumming et~al. 2012 in preparation;][for recent applications]{snl+12, akt+12}.
The method of calculation is the same as that of \citet[][]{bc09} for accreting neutron stars, but modified
to include the effect, of strong magnetic fields on the microphysics \citep{apm08}.

The model is essentially 1D, but the effect of the magnetic field on the heat transport is taken into
account by assuming a dipolar magnetic field and averaging over spherical shells \citep{py01, gh83}.
We use reasonable values for the neutron star mass ($1.4M_{\sun}$) and radius (12 km).
We then find the best set of parameters, the magnetic field ($B$) and the core temperature ($T_c$)
and the initial temperature profile as a function of density to explain the flux relaxation of a source
(see Cumming et~al. 2012 in preparation for more details).
 
We applied the model to match the light curve of CXOU~J164710.2$-$455216 after the 2006 outburst.
Figure~\ref{fig:coolcurve} shows sample cooling models that are qualitatively consistent with
the data.
The total energy and depth are robust, and correspond to $\sim1\times10^{44}$ ergs injected into
most of the outer crust. The profile of the heating must be such that the temperature profile is
close to being independent of density, with the exact slope depending on the $B$ that we assume. For
the models with $B=8\times10^{13}\ \rm G$ or $B=2\times10^{14}\ \rm G$, the temperature must
decrease slowly with density  $T\propto \rho^{-0.2}$,
whereas for $B=2\times 10^{15}\ \rm G$, an isothermal model reproduces the observed decay.

For $B=8\times 10^{13}\ {\rm G}$ or $B=2\times 10^{14}\ {\rm G}$, we find that the inner boundary of
the heated region has to be placed at a density $5\times 10^{11}\ {\rm g\ cm^{-3}}$, close to neutron
drip (at $\rho\approx 4\times 10^{11}\ {\rm g\ cm^{-3}}$). The fact that the inner boundary is so
close to the neutron drip density may indicate that the heating extends into the inner crust,
but that the temperature close to neutron drip where the neutrons are normal
(low critical temperature) remains low due to the neutron heat capacity.
This will be investigated further elsewhere (Cumming et al.~2012 in preparation).

The fact that the temperature profile is relatively independent of depth in the models is
interesting; if the energy was deposited as a fixed energy density
as assumed by \citet{let02} and corresponding to, for example, a fixed fraction of the magnetic energy density,
the temperature profile
would decrease quite steeply with increasing density ($T\propto \rho^{-0.6}$) and that would give a light
curve that drops too steeply for $t<100$ days, inconsistent with the data. Instead, the preferred energy
deposition profile is more like ``constant energy per gram'' (or equivalently per particle).
Crust breaking would give something close to this, because the maximum elastic energy that can be stored
in the crust is proportional to the shear modulus $\mu$, which is proportional to the pressure $\Pi$,
so that the energy density deposited by crust
breaking would be $\propto \mu \propto \Pi \propto \rho^{4/3}$, and the temperature profile be
$T \propto \rho^{0.3}$, increasing only slowly with density. Our model profiles sit somewhere between the
the fixed energy density and the maximal crust breaking cases, suggesting that crustal breaking might
have occurred only locally.

\begin{figure}
\centering
\vspace{0.1 in}
\begin{tabular}{c}
\vspace{0.1 in}
\includegraphics[width=2.25 in, angle=90]{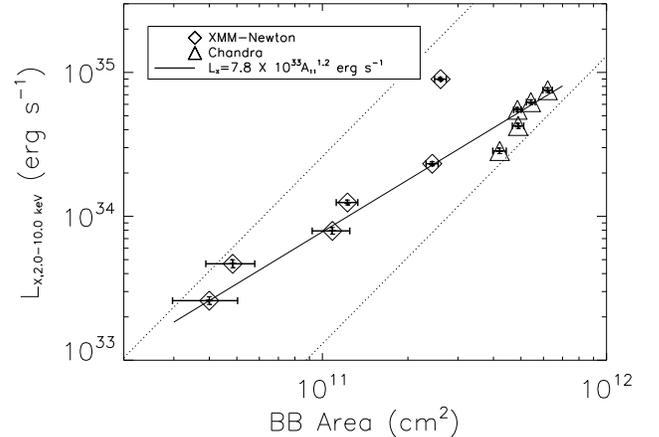} 
\end{tabular}
\vspace{-5.00mm}
\figcaption{Blackbody area versus 2--10 keV X-ray luminosity of the source after the 2006 outburst. The
solid line shows $L_X \propto A_{11}^{1.2}$, where $A_{11}$ is the blackbody area in units of
$10^{11}\ \rm cm^2$, and dashed lines show the simple models ($L_X \propto A_{11}^2$)
in Figure~3 of \citet{b10}.
\label{fig:LArea}
}
\end{figure}

Our models calculate the bolometric luminosity evolution, but we consider only
the 2--10 keV source flux since it is a more robust measurement than in the 0.5--10 keV, due to
the uncertainty in $N_{\rm H}$.
If we considered the bolometric flux, the shape of the flux evolution, and thus the model
parameters might change. It was difficult to investigate this with the spectral models we used,
since we are using a power-law model which has no lower-energy cutoff.
Instead, we attempted to determine bolometric fluxes by utilizing the RCS model fits.

Although the RCS model parameters were not well constrained (see Section~\ref{spectrumana}),
we were able to measure the fluxes well with this model.
The RCS fluxes agreed very well with those of the power-law
plus blackbody in the 2--10 keV band, and thus our crustal cooling models were able to fit the
RCS flux evolution without changing any parameter in this case.
The bolometric fluxes were higher and the discrepancy slowly increased ($F_{bolometric}/F_{2-10\ \rm{keV}}=1.7-2.4$)
as the flux decreased. The shape of the flux evolution changed in this case, so
one may expect the cooling model parameters to be modified. 
However, the quiescent flux level also increased, and thus
we had to adjust the constant offset of the crustal cooling models (core temperature).
The offset compensated for the change
of the shape, and no other parameters (except for the normalization which corresponds
to the total energy in the model) needed to be modified significantly.
For example, the core temperature of the model increased by $\sim40$\% ($1.5\times10^8\ \rm K$),
the total energy by a factor of $\sim2$ ($\sim4\times10^{44}$ ergs), and the inner boundary of the heated region had to be
placed at $\sim10$\% higher density ($\rho=5.5\times 10^{11}\ {\rm g\ cm^{-3}}$) for the low magnetic field model
($B=8\times 10^{13}\ {\rm G}$).

We note that the untwisting models \citep{b09} can produce diverse functional forms for the flux decay and
the blackbody area ($A_{bb}$) evolution, and may as well explain the evolution of source's flux and blackbody area
during the period over which the blackbody area decreased monotonically. For example, \citet{b09} shows a
sample cooling curve (see his Figure~7), and
\citet{b10} predicts that luminosity is quadratically proportional to the blackbody
area, for a model with constant discharge voltage (see his Figure~3).
However, we find $L_X \propto A_{bb}^{1.2}$ better describes the area versus luminosity relation of the source
after the 2006 outburst (see Fig.~\ref{fig:LArea}). Therefore, the simple model may not properly
describe the flux and the blackbody area evolution of the source. Investigating other models is beyond
the scope of this paper.

\begin{figure}[t]
\centering
\includegraphics[width=4.25 in]{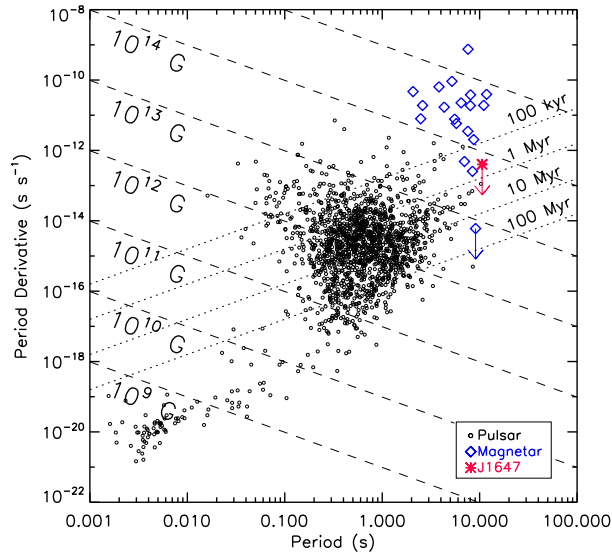}
\figcaption{Location of CXOU~J164710.2$-$455216 in the $P$--$\dot P$ diagram. It sits
well below the $B=10^{14}\ \rm G$ line. Pulsar data are taken from the ATNF pulsar
database and magnetars are from the McGill online magnetar catalog.
\label{fig:PPdotdiag}
}
\vspace{0.1 in}
\end{figure}

\subsection{Timing Properties}
\label{sec:timing}
\citet{icd+07} and \citet{wkg+11} measured the timing properties of the source with data covering
$\sim$100--200 days after the
2006 outburst. Their measurements together with the results of this work are summarized in Table~\ref{ta:timing}.
The measured periods at the outburst onset epoch agree well with one another. However, the period derivative we
measured is significantly smaller. This might be due to enhanced spin-down following a putative glitch
at the outburst epoch \citep{icd+07}, as has been seen in other sources
\citep[1RXS~J170849.0$-$400910, 1E~2259$+$586;][]{kg03,wkt+04}.
Indeed, \citet{wkg+11} noted that glitch recovery could
bias the magnitude of the spin-down rate upward for this source and that the spin-inferred dipolar magnetic field of
the source would be $\sim3.7\times10^{13}\ \rm G$ assuming that the glitch recovery was
completed by MJD 54148.
Although \citet{wkg+11} concluded that the presence of a glitch could not be conclusively demonstrated,
the small spin-down rate we measure supports the idea of
glitch recovery. If this is the case, our result implies that the spin-inferred dipolar magnetic
field of the source is less than $7\times10^{13}\ \rm G$, which is consistent with
$B=3.7\times10^{13}\ \rm G$ as inferred by \citet{wkg+11} and comparable to those of other apparently low
magnetic-field magnetars \citep[e.g., Swift~J1822.3$-$1606, 1E~2259$+$186;][]{lsk+11, rie+12, snl+12, gk02}
and high-$B$ rotation-powered pulsars \citep[e.g., PSR~J1718$-$3718;][]{nk11}.

With this measurement, CXOU~J164710.2$-$455216 can be added to
the growing list of  neutron stars displaying magnetar behavior but having relatively low magnetic fields
($< 10^{14}\ \rm G$), comparable to those of high-$B$ rotation-powered pulsars. Figure~\ref{fig:PPdotdiag}
shows the location of the
source in the $P\dot{P}$ diagram, where pulsar data are taken from the ATNF pulsar
database\footnote{http://www.atnf.csiro.au/research/pulsar/psrcat} \citep[][]{mht+05}
and magnetars are from the McGill online magnetar catalog.
Given similar magnetic fields, it appears that  some show magnetar-like behavior while others do not. This is
puzzling in the context of the magnetar model unless we assume higher order multipoles or strong internal
toroidal fields in some sources \citep[e.g., SGR~0418$+$5729; ][]{ret+10, ggo+11}.

We note that the dipolar magnetic field strengths measured using the standard dipolar braking relation
may not be correct since the formula includes many assumptions such as radius, mass and
magnetic inclination angle. For example, the inferred $B$-field strength can change by a factor of two due to
the inclination angle \citep{s06}.
Also the mass (for $M_{NS}=1.4\pm0.5M_{\sun}$) and the radius (for $R=12\pm2\ \rm km$) can
change it by $\sim15$\% and $\sim 30$\%, respectively. Although these uncertainties are fairly large, they are not
likely to explain the larger range (three orders of magnitude) of inferred magnetic field strengths now
observed in magnetars. 

\section{Conclusions}
\label{sec:concl}
Using archival data,
we have measured the spectral evolution of CXOU~J164710.2$-$455216 in the 2--10 keV band
for approximately 3 years since its 2006 outburst.
We see a clear correlation between spectral hardness and the flux; the spectrum softens as the flux declines.
Our timing analysis for data spanning approximately 2500 days shows that the spin-inferred dipolar magnetic
field of the source is less than $7\times10^{13}\ \rm G$. This is significantly lower than what has been
previously reported and supports the possibility of glitch recovery following the 2006 outburst.
This result adds to the growing list of relatively low-$B$ magnetars.
We find evidence of a second outburst based on a flux increase between MJD 55068 and 55832.
Finally, fitting the flux decay with a crustal cooling model suggests that the cooling trend of
CXOU~J164710.2$-$455216 after its 2006 outburst can be reproduced if energy was deposited
in the outer crust and the initial temperature
profile was relatively independent of depth.\\

We thank P.~M. Woods for useful discussions. V.M.K. acknowledges support
from a Killam Fellowship, an NSERC Discovery Grant, the FQRNT Centre de Recherche Astrophysique du Qu\'ebec,
an R. Howard Webster Foundation Fellowship from the Canadian Institute for Advanced
Research (CIFAR), the Canada Research Chairs Program and the Lorne Trottier Chair
in Astrophysics and Cosmology. A.C. is supported by an NSERC Discovery Grant and the
Canadian Institute for Advanced Research (CIFAR).

\end{document}